\documentclass{aa}
\usepackage{epsfig}
\usepackage{natbib}
\usepackage{amsmath}
\usepackage{amssymb}
\usepackage{enumerate}
\usepackage[toc,page]{appendix}

\begin{document}

\title{Modelling multiwavelength emissions from PSR B1259-63/LS 2883: Effects of the stellar disc on shock radiations}

\author{A. M. Chen\inst{1,2}
  \and J. Takata\inst{3}
  \and S. X. Yi\inst{2,4}
  \and Y. W. Yu\inst{1}
  \and K. S. Cheng\inst{2}}

\institute{
Institute of Astrophysics, Central China Normal University, Wuhan 430079, China
\and Department of Physics, The University of Hong Kong, Pokfulam Road, Hong Kong
\and School of Physics, Huazhong University of Science and Technology, Wuhan 430074, China
\and Department of Astrophysics, Radboud University Nijmegen, P.O. Box 9010, NL-6500 GL Nijmegen, The Netherlands}

\date{Received xxx / Accepted xxx}

\abstract{
PSR B1259-63/LS 2883 is an elliptical pulsar/Be star binary that emits broadband emissions from radio to TeV $\gamma$-rays. The massive star possesses an equatorial disc that is inclined with the orbital plane of the pulsar. Non-thermal emission from the system is believed to be produced by pulsar wind shock and double-peak profiles in the X-ray, and TeV $\gamma$-ray light curves are related to the phases of the pulsar passing through the disc region of the star. In this paper, we investigate the interactions between the pulsar wind and stellar outflows, especially with the presence of the disc, and present a multiwavelength modelling of the emission from this system. We show that the double-peak profiles of X-ray and TeV $\gamma$-ray light curves are caused by the enhancements of the magnetic field and soft photons at the shock during the disc passages. As the pulsar is passing through the equatorial disc, the additional pressure of the disc pushes the shock surface closer to the pulsar, which causes the enhancement of magnetic field in the shock, and thus increases the synchrotron luminosity. The TeV $\gamma$-rays due to the inverse-Compton (IC) scattering of shocked electrons with seed photons from the star are expected to peak around periastron, which is inconsistent with observations. However, the shock heating of the stellar disc could provide additional seed photons for IC scattering during the disc passages, and thus produces the double-peak profiles as observed in the TeV $\gamma$-ray light curve. Our model can possibly be examined and applied to other similar gamma-ray binaries, such as PSR J2032+4127/MT91 213, HESS J0632+057, and LS I+61$^{\circ}$303.

\keywords{binaries: close -- Gamma rays: stars -- X-rays: binaries -- pulsars: individual(PSR B1259-63)}
}

\authorrunning{Chen et al.}
\titlerunning{Modelling the emissions from PSR B1259-63/LS 2883}

\maketitle

\section{Introduction}
High-mass gamma-ray binaries are a small class of binary systems composed of a compact object which is orbiting around a massive main sequence star and radiate $\gamma$-rays with the emission peaking at energies above 1 MeV (see Dubus 2013 and Paredes \& Bordas 2019 for reviews on gamma-ray binaries). The compact object is widely believed to be a rotating pulsar, although it still needs to be proved for most binaries. PSR B1259-63/LS 2883 is the first of these binaries whose compact object has been identified as a radio pulsar. The pulsed radio emission indicates the spin period of the pulsar is $P=47.76\ \rm{ms}$ with a spin-down power of $L_{\rm sd}\simeq8\times10^{35}\ \rm{erg/s}$ (Johnston et al. 1992). The pulsar is moving in a highly elliptical orbit ($e\simeq0.87$) around the companion star with an orbital period of $1237$ days, and the system is about $2.60$ kpc away from the Earth (Johnston et al. 1992; Negueruela et al. 2011; Shannon et al. 2014; Miller-Jones et al. 2018).

After its first discovery in 1992, PSR B1259-63/LS 2883 has been widely detected from radio to very high energy $\gamma$-rays (e.g. radio: Johnston et al. 1992, 1994, 1996, 1999, 2001, 2005; Manchester et al. 1995; Wex et al. 1998; Connors et al. 2002; Wang et al. 2004;  Mold$\acute{\rm o}$n et al. 2011; Shannon et al. 2014; Miller-Jones et al. 2018; optical: Johnston et al. 1994; Negueruela et al. 2011; van Soelen et al. 2011, 2012, 2016; X-ray: Cominsky et al. 1994; King \& Cominsky 1994; Kaspi et al. 1995; Hirayama et al. 1999; Shaw et al. 2004; Chernyakova et al. 2006, 2009; Uchiyama et al. 2009; Pavlov et al. 2011, 2015; Kargaltsev et al. 2014; GeV $\gamma$-ray: Abdo et al. 2011; Tam et al. 2011, 2015, 2018; Caliandro et al. 2015; Xing et al. 2016; Johnson et al. 2018; Chang et al. 2018; TeV $\gamma$-ray: Aharonian et al. 2005, 2009; Abramowski et al. 2013; Romoli et al. 2015, 2017). 
Two extended multiwavelength observational campaigns were performed on this system during the 2010 and 2014 periastron passages (Chernyakova et al. 2014, 2015). The multiwavelength emission varied with the orbital motion of the pulsar and did not show significant super-orbital modulations. In particular, the un-pulsed radio, X-ray, and TeV $\gamma$-ray light curves are similar for previous periastron passages and they are characterized by two asymmetrical peak profiles around periastron.
When the pulsar is moving around apastron, the radio emission comprises only the pulsed component. As the pulsar is approaching periastron, the pulsed radio flux decreases and finally disappears during $\sim T_0\pm15$ days ($T_0$ is the time when the pulsar is at periastron), while the un-pulsed radio flux increases and reaches its maximums at $\sim T_0-10$ days and $\sim T_0+20$ days (e.g. Chernyakova et al. 2014, 2015).
The X-ray emission is detected throughout the entire orbit and exhibits a similar behaviour to the un-pulsed radio light curve. About 25 days before periastron, there is a rapid increase in the X-ray flux reaching its maximum at $\sim T_0-15$ days. Subsequently, the X-ray emission decreases gradually to a minimum around periastron and then it is followed by a second peak around $\sim T_0+20$ days (e.g. see Chernyakova et al. 2014, 2015).
Although the detections by the High Energy Stereoscopic System (\textit{H.E.S.S.}) of PSR B1259-63/LS 2883 are not adequate to reach final conclusions, the combined TeV $\gamma$-ray light curve with all available observational data shows double asymmetrical peaks around periastron, which is similar to the X-ray and un-pulsed radio light curves (Sushch \& B$\ddot{\rm o}$ttcher 2014; Romoli et al. 2017).
The overall double-peak profiles as observed in the radio, X-ray, and TeV $\gamma$-ray light curves are generally believed to be attributed to the phases of the pulsar passing through the dense disc regions of the star (e.g. Chernyakova et al. 2014, 2015).
In addition, the GeV $\gamma$-rays from PSR B1259-63/LS 2883 was detected by \textit{Fermi}/LAT during the 2010, 2014, and 2017 periastron passages (e.g. Johnson et al. 2018; Tam et al. 2018; Chang et al. 2018).
The 2010 and 2014 analysis results showed that
only upper limits or a very low flux was detected when the pulsar is close to periastron. Afterwards, it was accompanied by an enhancement of GeV flux starting at 30-32 days after periastron and the flare lasted for several weeks. Strangely, in 2017, the flare started at $\sim T_0+40$ days which is about 10 days later than those in 2010 and 2014, and showed notable spectral curvature and rapid variability in the light curves. Also, the $\gamma$-ray luminosity in 2017 is brighter than previous flares with the peak flux levels even exceeding the spin-down luminosity of the pulsar (e.g. Johnson et al. 2018).
The origin of the distinctive GeV emission is still unclear.

In the leptonic model of pulsar-powered binaries, the interactions between the relativistic wind of the pulsar and the outflows from the companion star form a terminal shock in which the ram pressures of two winds are in balance. The electron and positron pairs (hereafter electrons) from the pulsar wind are accelerated to ultra-relativistic speeds by the terminal shock and emit broadband non-thermal emission via synchrotron radiation and IC scattering (Tavani \& Arons 1997; Kirk, Ball \& Skj\ae raasen 1999; Dubus 2006b). Under the pulsar wind shock model, the X-ray and TeV $\gamma$-ray emissions are expected to peak around periastron owing to a higher magnetic field and a higher density of stellar photons at the shock along the entire orbit\footnote{Considered the anisotropic nature of IC, the $\gamma$-ray light curve peaks a few days before periastron due to a favourable scattering angle.}. In the revised shock scenario, some other effects are considered: for example, the Doppler-boosting effect enhances the shock emissions around the inferior conjunction due to a more favourable orientation of the shocked flows (Dubus, Cerutti, \& Henri 2010); the pair creation process suppresses the TeV $\gamma$-rays around periastron owing to a denser soft photon density and a more efficient collision angle between the $\gamma$-rays and stellar photons (Dubus 2006a; Sushch \& van Soelen 2017); moreover, the non-radiative cooling of shocked electrons (Khangulyan et al. 2007; Kerschhaggl 2011) and the anisotropy of pulsar wind (Kong et al. 2012) are introduced to explain the complex emission behaviours of PSR B1259-63/LS 2883.
Besides the leptonic model mentioned above, it was suggested that the non-thermal emissions can also be produced by the hadronic process (e.g. Kawachi et al. 2004; Neronov \& Chernyakova 2007). The protons from the pulsar wind would be accelerated to relativistic speeds by the terminal shock and then converted into pions after the collisions with the protons from the stellar outflows. The high energy $\gamma$-rays are produced by the decays of neutral pions and the secondary particles generated by the decays of charged pions also contributes to the observed emissions.

One of the most important features of this binary system is that the companion star has an equatorial disc which is inclined with the orbital plane, so the pulsar passes through the disc region twice in one orbital period (Melatos, Johnston $\&$ Melrose 1995; Chernyakova et al. 2006).
The presence of the stellar disc on shock radiation was firstly investigated in Tavani \& Arons (1997). In their model, the orbital variation of the X-rays is caused by the changes of the magnetic field in the shock and the increase of energy loss of shocked electrons due to IC process when the pulsar is moving around periastron.
Sierpowska-Bartosik \& Bednarek (2008) studied the effect of an aspherical stellar outflows on the shock structure and found that the geometry of the shock would change significantly as the pulsar moves from the dilute polar wind to the dense stellar disc region. These authors considered the complex shock structure due to the stellar outflows and the anisotropic photon field on the IC scattering with pair cascades to explain the TeV emissions.
Besides the optical photons from the companion star, the stellar disc also produces an infrared (IR) excess which can be up-scattered to high energy $\gamma$-rays (van Soelen $\&$ Meintjes 2011). van Soelen et al. (2012) found that the $\gamma$-ray emission produced by the IC process of shocked electrons with the disc photons is expected to peak around periastron instead of the disc passages. This is because the IR excess is mainly generated by the inner part of the disc, and its maximum contribution on IC scattering is around periastron when the pulsar is closest to the brighter disc region (van Soelen et al. 2012).
In addition to its contribution on IC, the IR emission from the stellar disc also provides additional soft photon for pair cascades and $\gamma$-ray absorption. In Sushch \& B$\ddot{\rm o}$ttcher (2014), they studied the pair cascades process of $\gamma$-rays assuming a constant width of the disc and a constant energy density inside the disc. These authors suggested that the cascade emission may explain the relatively low GeV flux around periastron, but cannot produce the observed GeV flare.
A detailed investigation on $\gamma$-ray absorption of PSR B1259-63/LS 2883 was presented in Sushch \& van Soelen (2017). They found that the $\gamma$-ray absorption due to the disc photons produces a $\approx14\%$ decrease in the TeV flux and the maxima absorption occurs about four days before periastron and the total absorption due to the stellar and disc photons produces a maximum decrease with $\approx52\% $ around periastron.
Okazaki et al. (2011) used three-dimensional smooth particle hydrodynamical (SPH) simulations to investigate the tidal and wind interactions with the presence of the stellar disc of PSR B1259-63/LS 2883. The simulations indicated that the strong pulsar wind strips off the outer part of the disc, and truncates the disc at a distance much smaller than
the orbit of the pulsar. Based on the work of Okazaki et al. (2011), Takata et al. (2012) studied the influence of the stellar disc on the non-thermal emission from the binary system. The double-peak profiles in the X-ray light curve was attributed to a significant increase of particle luminosity during the disc passages. The predicted TeV light curve, however, showed the maximum flux at periastron which is inconsistent with the observations.

Although the double-peak profiles of the un-pulsed radio, X-ray, and TeV $\gamma$-rays in PSR B1259-63/LS 2883 are generally thought to be connected to the phases of the pulsar passes through the stellar disc, the detail mechanisms for the enhancements of the shock emissions are still unclear. In this paper, we investigate the interaction between the relativistic wind from the pulsar and the aspheric stellar outflows from the Be star, especially with the presence of its equatorial disc. We study how the disc affects the structure of the terminal shock and the radiation process, and discuss the formation of the double-peak structures as observed in the X-ray and TeV $\gamma$-ray light curves.
The outline of our paper is as follows: we describe the wind interaction in Section 2.1; the electron distribution and the shock radiations are presented in Section 2.2. Then we present the theoretical results with comparison of observations in Section 3.1, and discuss the effects of the different disc parameters on the X-ray and TeV $\gamma$-ray light curves in Section 3.2. Finally, a brief discussion and conclusion are provided in Section 4.

\section{Model description}

\begin{figure*}
\centering
\resizebox{0.6\hsize}{!}{\includegraphics{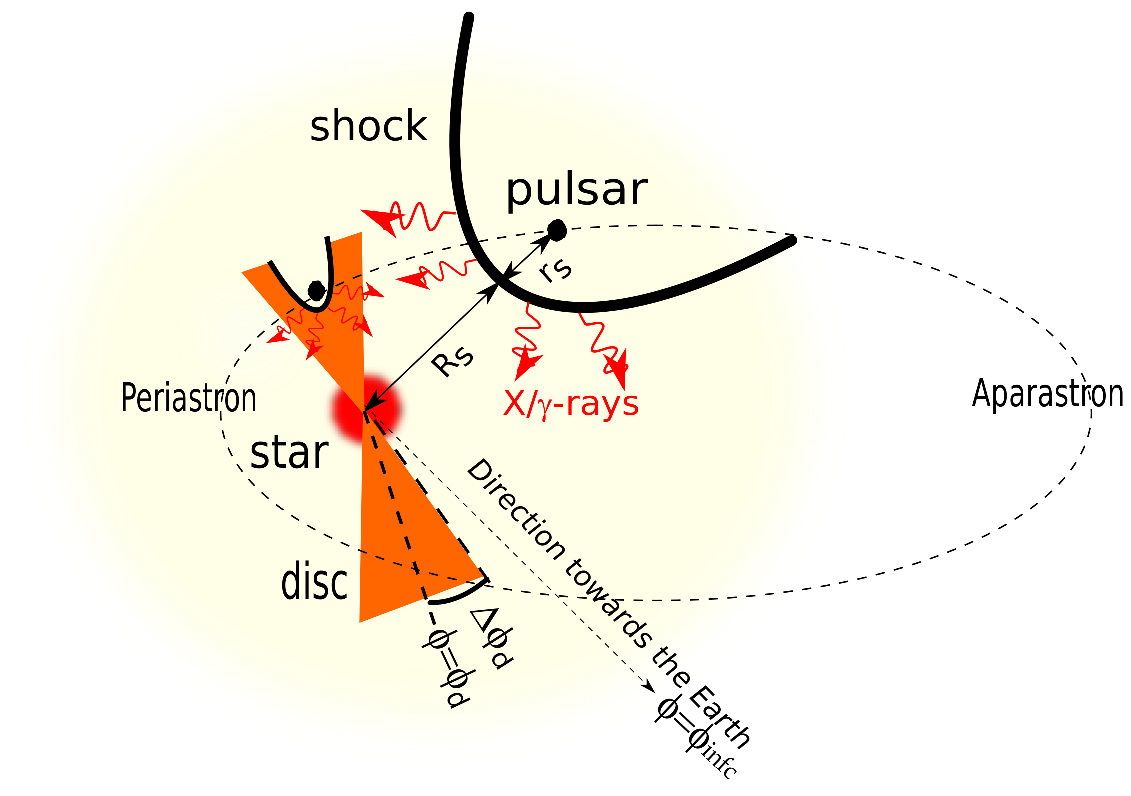}}
\caption{
Illustration of the binary geometry. The interaction between the pulsar wind and stellar outflows forms a terminal shock with a bow-shape geometry. The electrons from the pulsar wind are accelerated to very high energies by the shock, and then emit the keV X-rays and TeV $\gamma$-rays through synchrotron radiation and IC scattering, respectively. The shadow area is the stellar disc of the Be star projected on the orbital plane. The direction towards the Earth is indicated in the plot.}\label{fig:binary}
\end{figure*}

In Fig.\ \ref{fig:binary}, we illustrate the geometry of the binary system as discussed in this paper. The pulsar wind is stopped by the stellar outflows and forms a terminal shock with a bow-shape geometry. As the pulsar is passing through the disc, the pulsar wind shock is compressed because of the additional pressure from the stellar disc. The electrons from the pulsar wind are accelerated to very high energies by the shock and emit keV X-rays and TeV $\gamma$-rays through synchrotron radiation and IC scattering, respectively.

\subsection{Interactions between the pulsar wind and stellar outflows}

The star, LS 2883, is a massive Be star and the stellar outflows are characterized by a dilute polar wind with an equatorial disc (Waters et al. 1988). The mass-loss rate $\dot{M}$ and velocity of the stellar outflows $v_{\rm w}$ are distinct in the equatorial and polar regions. For a typical polar wind, we have $\dot{M}\sim10^{-8}\ M_{\odot}/{\rm yr}$ with $v_{\rm w}\sim10^8\ {\rm cm/s}$, while the equatorial disc is much denser than the polar wind with a slower speed. To get the dynamic pressure of the stellar outflows, we need to know the exact mass-loss rates and velocity profiles of the polar wind and equatorial disc, which could be complex in the case of Be stars (Khangulyan et al. 2011). We apply a simple model on the outflow structure of Be stars as explored in Petropoulou et al. (2018) and write the dynamic pressure of the stellar outflows of LS 2883 in the coordinate of the star as
\begin{equation}\label{Pw}
  P_{\rm w}(R,\theta)=p_0R^{-n}(1+G\mid\cos\theta\mid^m),
\end{equation}
where $R$ is the distance from the centre of the Be star, $\theta$ is the polar angle measured from the stellar equator, and $p_0$ is the ram pressure of the polar wind. The radial dependence of the wind pressure adopted in this case is $n=2$ as suggested by Petropoulou et al. (2018). The parameter $G$ is the equator-to-polar ram pressure contrast ($G=0$ corresponding to case without the disc), and the confinement of the equatorial disc is determined by the index $m$ (Ignace $\&$ Brimeyer 2006), i.e.
\begin{equation}\label{dtheta}
  \Delta\theta_{\rm d}=\cos^{-1}(G^{-1/m}),
\end{equation}
where $\Delta\theta_{\rm d}$ is the half-opening angle of the disc given in the coordinate of the star. Within this parameterization, a smaller value of $m$ means that the disc opening angle would be larger.

The equatorial disc of LS 2883 is thought to be inclined with the orbital plane (Melatos, Johnston $\&$ Melrose 1995; Chernyakova et al. 2006). The inclination angle of the disc is still unknown yet. Melatos, Johnston $\&$ Melrose (1995) suggested that the inclination angle is about $i_{\rm d}\sim10^{\circ}-40^{\circ}$ to fit the observed variations in dispersion measure and rotation measure of PSR B1259-63, while other values, such as $45^{\circ}$ and $90^{\circ}$, were also considered in theoretical models (Okazaki et al. 2011; Takata et al. 2012; van Soelen et al. 2012). Since the pulsar is moving on the orbital plane, it is more convenient to express the ram pressure of stellar outflows as a function of the true anomaly angle $\phi$ instead of the polar angle $\theta$ measured from the stellar equator. According to the rotation of the coordinate system, we have\begin{equation}\label{cs}
  \sin{\theta} =-\sin{i_{\rm d}}\cdot\sin(\phi-\phi_{\rm d}),
\end{equation}
where $\phi_{\rm d}$ is the true anomaly of the midplane of the disc. For a perpendicular disc (i.e. $i_{\rm d}=90^{\circ}$), the above equation can be simply written as $\theta=-(\phi-\phi_{\rm d})$. Substituting Eq.(\ref{cs}) into Eq.(\ref{Pw}), we can get the ram pressure of stellar outflows on the orbital plane. Similarly, the half-opening angle of the disc projected on the orbital plane is given by
\begin{equation}\label{dphi}
  \Delta\phi_{\rm d}=\sin^{-1}\left(\frac{\sin\Delta\theta_{\rm d}}{\sin{i_{\rm d}}}\right).
\end{equation}
With a fixed value of $\Delta\theta_{\rm d}$ as given in Eq.(\ref{dtheta}), if the inclination angle $i_{\rm d}$ is smaller,  the disc region on the orbital plane would be larger, and it takes a longer time for the pulsar to pass through the disc region.

The position and geometry of the terminal shock are determined by the dynamic balance of two winds. We define the momentum flux ratio of the pulsar wind to stellar outflows as
\begin{equation}\label{eta}
  \eta(\phi)=\frac{L_{\rm sd}/c}{4\pi p_0[1+G\mid\cos\theta(\phi)\mid^m]},
\end{equation}
then the shock distance from the pulsar can be written as (Canto et al. 1996)\begin{equation}\label{rs}
  r_{\rm s}(\phi)=d \frac{\eta^{1/2}}{1+\eta^{1/2}},
\end{equation}
where $d$ is the binary separation. The terminal shock has a bow-shape geometry and the half-opening angle can be estimated as (Eichler \& Usov 1993)
\begin{equation}\label{thetas}
\theta_{\textrm{s}}(\phi)=2.1(1-\eta^{2/5}/4)\eta^{1/3}.
\end{equation}
Because of the anisotropy of the stellar outflows (i.e. $G\neq0$), the momentum flux ratio $\eta$ is expected to vary along the orbital period, so the shock distance and opening angle of the terminal shock also varies with orbital phase. In particular, when the pulsar is moving from the dilute wind region to the dense disc environment, the shock surface is pushed closer to the pulsar because of the additional pressure of the disc, and the opening angle of the shock also becomes smaller.

After the shock distance is determined, we can obtain the magnetic field in the shock from the magnetohydrodynamic jump conditions (Kennel \& Coroniti 1984a,b) as follows:
\begin{equation}\label{B}
B=\sqrt{\frac{L_{\rm sd}\sigma}{r_{\rm s}^2c(1+\sigma)}\left(1+\frac{1}{u^2}\right)},
\end{equation}
\begin{equation}\label{u}
u^2=
\frac{8\sigma^2+10\sigma+1}{16(\sigma+1)}+
\frac{[64\sigma^2(\sigma+1)^2+20\sigma(\sigma+1)+1]^{1/2}}{16(\sigma+1)},
\end{equation}
where $\sigma$ is the magnetization parameter of pulsar wind at the shock. According to the previous studies of pulsar wind nebula, the wind should be kinetic dominated at a much larger distance from the light cylinder of the pulsar, and it means that the value of $\sigma$ at the shock cannot be very large (e.g. Kennel \& Coroniti 1984a,b). Sierpowska-Bartosik \& Bednarek (2008) suggested the wind magnetization parameter should be $\sigma\sim10^{-2}-10^{-4}$ for PSR B1259-63/LS 2883. So we adopt $\sigma=10^{-3}$ in the calculation based on the fitting of the X-ray data. With a constant value of $\sigma$, we have $B\propto r_{\rm s}^{-1}$, which means that the strength of magnetic field in the shock becomes larger for a smaller shock distance from the pulsar. As we can imagine when the pulsar passes through the disc region, the additional pressure of the disc pushes the shock closer to the pulsar, so the strength of magnetic field in the shock region is enhanced during the disc passages.

The energy density of radiation field from the companion star in the shock is given by
\begin{equation}\label{ustar}
  u_{\rm star}=\frac{L_{\rm star}}{4\pi R_{\rm s}^2c},
\end{equation}
where $L_{\rm star}$ is the luminosity of the star, and $R_{\rm s}=d-r_{\rm s}$ is the shock distance from the centre of the star. Kirk, Ball \& Skj\ae raasen (1999) investigated the IC scattering in the shock with the seed photons originating from the companion star, and found that the predicted TeV $\gamma$-rays showed the highest flux around periastron where the energy density of photon field is highest. It means that the observed enhancements in the TeV flux is unlikely to be caused by the up-scattering of stellar photons.
Alternatively, van Soelen \& Meintjes (2011) and van Soelen et al. (2012) studied the influence of the IR excess from the stellar disc of LS 2883 on the shock radiation, and the predicted $\gamma$-ray light curve shows that the maxima contributions due to IC scattering with the disc photons is also around periastron instead of the disc passages. Because the IR photons are mainly produced by the inner part of the disc, its maximum contribution to IC scattering occurs around periastron, where the pulsar is closest to the brightest region of the disc. At the location where the pulsar crosses the disc ($\sim31R_{\star}$ and $\sim40R_{\star}$ from the centre of the disc during the pre- and post-periastron disc passages), the local intensity of disc emission is much lower with longer wavelengths, and its contribution to the production of TeV $\gamma$-rays is very small (van Soelen et al. 2012). It means that the IC scattering in the shock with the stellar photons or the IR emissions generated by the disc itself cannot explain the observed double-peak profiles of the TeV light curves, and additional seed photon components are required under the IC scattering scenario.

Another possible component for IC scattering is the optical/IR emission produced by the shock heating of the stellar disc as discussed in van Soelen et al. (2012). The heating process during the disc passages could increase the disc intensity and thus increase the $\gamma$-ray luminosity by IC scattering.
The total energy for the shock heating of disc matter can be obtained by estimating the kinetic energy of the stellar disc being converted into the internal energy of the shocked disc matter. With the dynamic balance between the stellar disc and pulsar wind $\rho_{\rm w}v_{\rm w}^2\simeq L_{\rm sd}/4\pi r_{\rm s}^2c$, the energy density of the shocked disc matter can be written equally in the form of the spin-down luminosity of the pulsar:
\begin{equation}\label{udisc}
  u_{\rm disc}\simeq\xi\frac{L_{\rm sd}}{4\pi r_{\rm s}^2c},\ \rm{for}\ \phi\in[\phi_{\rm d}\pm\Delta\phi_{\rm d}]\&[(\phi_{\rm d}+\pi)\pm\Delta\phi_{\rm d}]
\end{equation}
where $\xi$ is the shock heating efficiency.
Initially, the emerging radiation from the shocked disc region is dominated by the thermal bremsstrahlung photons with the temperature of $kT_{\rm sh}=3\mu m_{\rm p}v_{\rm w}^2/16\sim12\ {\rm eV}$, where $\mu\simeq0.62$ is the average atomic weight, $m_{\rm p}$ is the mass of proton, and $v_{\rm w}\simeq10^7\ \rm{cm/s}$ is the velocity of the stellar disc (Zabalza et al. 2011a).
As the pulsar is passing through the dense disc environment, the strong wind of the pulsar sweeps up the disc matter into a dense shell accumulated at the shock front, and most of the  bremsstrahlung photons are absorbed by the shell. We assume that the thermalization is quickly established, then the emerging emission from the shock heating disc matter would be in the form of black-body photons. The temperature of the shock heating of disc matter can be estimated by
\begin{equation}\label{Tdisc}
  T_{\rm disc}\simeq\left(\frac{\xi L_{\rm sd}}{4\pi r_{\rm s}^2\sigma_{\rm SB}}\right)^{1/4}\\
  \sim1\times10^4{\rm K}\left(\frac{\xi L_{\rm sd}}{10^{35}{\rm erg/s}}\right)^{1/4}\left(\frac{r_{\rm s}}{10^{12}{\rm cm}}\right)^{-1/2},
\end{equation}
where $\sigma_{\rm SB}$ is the Stefan-Boltzmann constant. As expected, the optical/IR photons produced by the shock heating of the disc could provide additional soft photons for IC scattering during the disc passages.


\begin{figure*}[th]
\centerline{
\includegraphics[width=14cm,height=8.0cm]{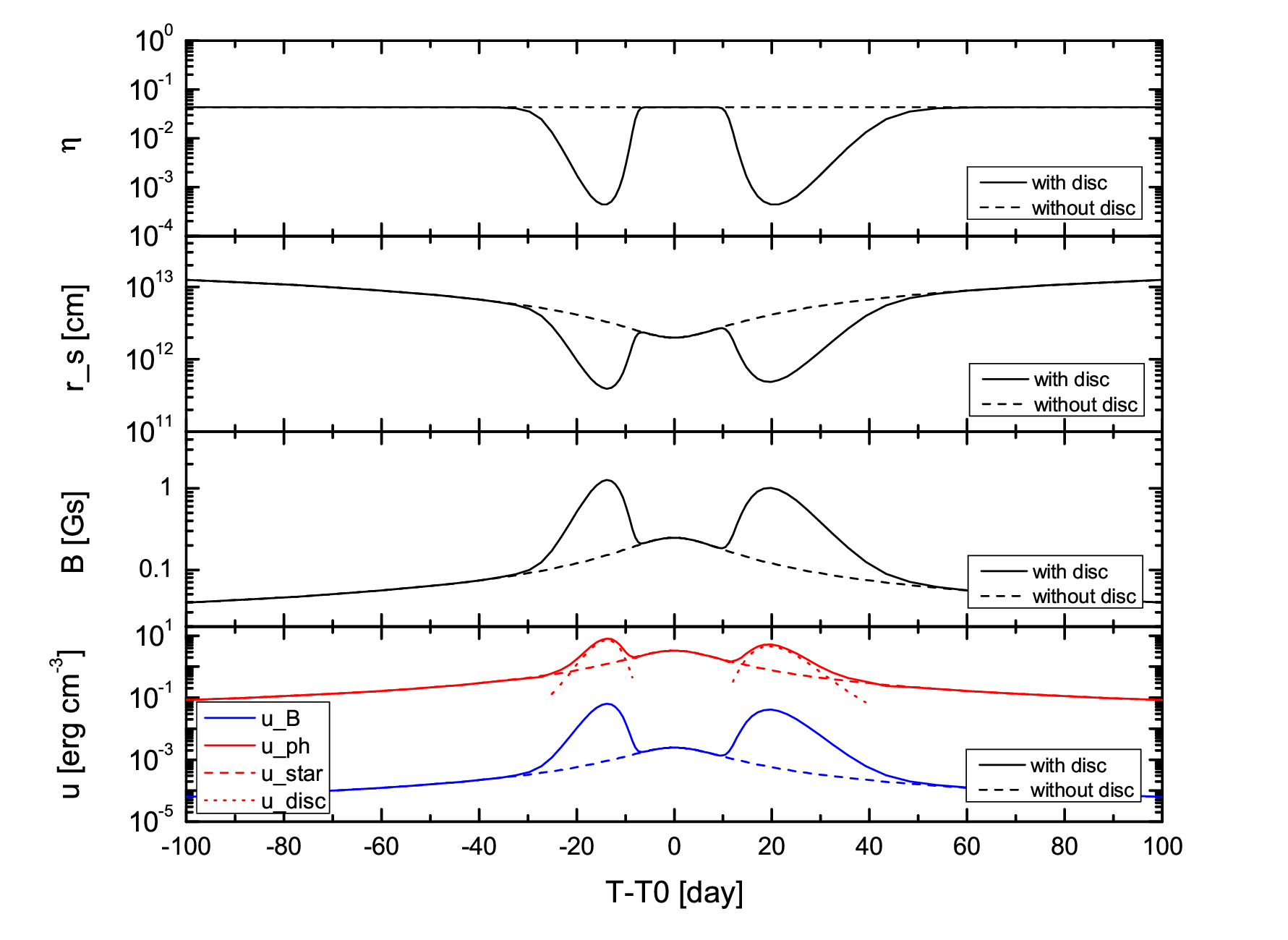}}
\caption{Orbital variations of shock parameters. From top to bottom: the momentum flux ratio of two winds $\eta$, the shock distance from the pulsar $r_{\rm s}$, the magnetic field strength $B(r_{\rm s})$, and the energy densities of the magnetic field $u_{\rm B}$ (blue lines) and the photon field $u_{\rm ph}$ (red lines) at the shock. The dashed lines correspond to the case without the presence of the stellar disc ($G=0$), while the solid lines show the effect of the disc on the shock parameters ($G=100, m=100$). In the bottom panel, the red dashed line shows the energy density of stellar photons $u_{\rm star}$, while the dotted line shows the energy density of photon field due to shock heating of the disc $u_{\rm disc}$. The red solid shows the sum of both components $u_{\rm ph}=u_{\rm star}+u_{\rm disc}$.}  \label{fig:shock}
\end{figure*}

In Fig.\ \ref{fig:shock}, we show the orbital variations of the momentum flux ratio of the pulsar wind to stellar outflows $\eta$, the shock distance from the pulsar $r_{\rm s}$, the strength of the magnetic field $B(r_{\rm s})$, and the energy densities of the magnetic field $u_B=B^2/8\pi$ with the photon field $u_{\rm ph}=u_{\rm star}+u_{\rm disc}$ at the shock. The dashed lines correspond to the case without the presence of the stellar disc, while the solid lines show the effects of the disc on the shock parameters. The model parameters are given in Table 1, which are chosen by modelling the observational data. In particular, the ram pressure of polar wind adopted in this work is $5\times 10^{25}\ \rm{g\cdot cm\cdot s^{-2}}$ to be consistent with typical values of the mass-loss rate and the velocity of the polar wind from a Be star ($\dot{M}\sim10^{-8}\ M_{\odot}/{\rm yr}$ and $v_{\rm w}\sim10^8\ {\rm cm/s}$). Given the uncertainty of the inclination angle of the disc, we adopt a modest value of $i_{\rm d}\sim45^{\circ}$, and other two disc parameters are fixed as $G=100$ and $m=100$.
How the disc parameters may affect the shock radiations is discussed in later sections. With the disc parameters given above, the half-opening angle of the disc project on the orbital plane is $\Delta\phi_{\rm d}\simeq17.25^{\circ}$. This value is slightly smaller than the value of $18.5^{\circ}$ given by Chernyakova et al. (2006), which was obtained by using the Gaussian function fitting of the observational data.
As we can see in the figure, considering the case of the polar wind without the disc, the momentum flux ratio $\eta$ is constant along the whole orbit, so the shock distance $r_{\rm s}$ is smallest at periastron. Since $u_{\rm B}\propto r_{\rm s}^{-2}\propto d^{-2}$ and $u_{\rm star}\propto R_{\rm s}^{-2}\propto d^{-2}$, the energy densities of the magnetic field and stellar photons in the shock are expected to show maximum at periastron.
However, with the presence of the stellar disc, the momentum flux ratio of two winds and shock distance from the pulsar show two dips during the disc passages due to the increase of the dynamic pressure of stellar outflows. When the pulsar is moving from the dilute wind region to the dense disc environment, the additional pressure of the equatorial disc pushes the shock surface closer to the pulsar. Therefore, the magnetic field and the magnetic energy density in the terminal shock is enhanced. Also, considering the shock heating process of the disc matter, the energy density of photon field is increased during the disc passages. In the calculations, we adopted a value for the heating efficiency of $\xi\simeq12.5\%$, and simply ignore the initial energy density of photon field inside the disc, since its initial intensity is small (van Soelen et al. 2012).

\begin{table*}[t]
\caption{Parameters of PSR B1259-63/LS 2883. \label{table}}
\begin{tabular}{l c c c}
\hline
\textbf{Parameter} &  \textbf{Symbol} & \textbf{Value}  & \textbf{Reference}\\
\hline
{\it System parameters} &  &  &\\
eccentricity$^\dag$  &  $e$  & 0.86987970  & 1\\
orbital period$^\dag$   &$P_{\textrm{orb}}$  & 1236.724526 \textrm{days}& 1\\
distance from Earth$^\dag$   &  $d_{\rm L}$   & 2.60 \textrm{kpc} & 2\\
inclination angle of the orbit$^\dag$  &$i$   & $26^{\circ}$ & 1\\
longitude of periastron$^\dag$  & $\omega$ & $138.665013^{\circ}$  & 2\\

\hline
{\it Pulsar and pulsar wind} &   &  & \\
spin-down power$^\dag$  & $L_{\textrm{sd}}$   &  $8.2\times 10^{35}\ {\rm erg\cdot s^{-1}}$   &3 \\
rotation period$^\dag$  &  $P$  & 47.7625 {\rm ms}  & 1\\
Lorentz factor of pulsar wind$^\ddag$ & $\gamma_\textrm{w}$  & $5\times10^5$  & -\\
magnetization of pulsar wind$^\ddag$  & $\sigma$  & $10^{-3}$  & -\\

\hline
{\it Star and stellar outflows}  &  &  & \\
effective radius of star$^\dag$  & $R_\star$  & $9.2 R_\odot$ & 4\\
effective temperature of star$^\dag$ & $T_\star$  & $3.02\times10^4\ \textrm{K}$  & 5\\
ram pressure of polar wind$^\ddag$  & $p_0$  & $5\times 10^{25}\ \rm{g\cdot cm\cdot s^{-2}}$  & -\\
radial profile of the wind$^\ddag$  & $n$  & $2$  & -\\
midplane of the disc $^\ddag$  & $\phi_{\rm d}$  &$97.5^{\circ}$  & -\\
inclination angle of the disc$^\ddag$  & $i_{\rm d}$  &$45^{\circ}$  & -\\
disc-wind pressure contrast$^\ddag$  &$G$&   $100$ & -\\
confinement of disc$^\ddag$  &$m$   &$100$  & -\\

\hline
{\it Terminal shock} &  &  & \\
particle distribution index$^\ddag$    & $p$   & $2.4$   & -\\
shock heating efficiency$^\ddag$    & $\xi$   & $12.5\%$   & -\\
acceleration efficiency$^\ddag$    & $\zeta$   & $0.3$   & -\\
velocity of shocked flow$^\ddag$   & $\beta$   & $1/3$  & -\\
\hline
\end{tabular}
\\
{{
\textbf{Notes.}\footnotesize $^\dag$ Observational parameters. \footnotesize $^\ddag$ Model parameters. The values adopted in this table are chosen by modelling the observational data.
\\
\textbf{Reference.} (1) Shannon et al. (2014); (2) Miller-Jones et al. (2018); (3) Manchester et al. (1995); (4) Negueruela et al. (2011); (5) Chernyakova et al. (2014).
}}
\end{table*}

\subsection{Electron distribution and shock radiation}
The electrons from the cold pulsar wind are accelerated to ultra-relativistic speeds via Fermi mechanism or reconnection process in the terminal shock. We assume that the injection rate of electrons in the shock follows a power-law distribution,
\begin{equation}\label{Qe}
  Q(\gamma,t)=K\gamma^{-p}\quad {\rm for}\quad \gamma_{\rm min}<\gamma<\gamma_{\rm max},
\end{equation}
where $K=(1-p)/(\gamma_{\max}^{1-p}-\gamma_{\min}^{1-p})$ is the normalization factor. The minimum Lorentz factor of the shocked electrons is determined by the Lorentz factor of the pulsar wind at the pre-shock region $\gamma_{\rm w}$, which is given by $\gamma_{\min}\simeq\gamma_{\rm w}(p-2)/(p-1)$. Since the pulsar wind is dominated by the kinetic energy of particles at a much larger distance from the pulsar, the value of $\gamma_{\rm w}$ can be estimated as $\gamma_{\rm w}\sim\sigma_{\rm L}\gamma_{\rm L}$, where $\sigma_{\rm L}$ and $\gamma_{\rm L}$ are the magnetization parameter and the Lorentz factor of electrons at the light cylinder, respectively. With the typical values of $\sigma_{\rm L}\sim10^3$ and $\gamma_{\rm L}\sim10^2-10^3$, we adopt $\gamma_{\rm w}\sim5\times10^5$ in the calculations (Takata et al. 2012). The maximum Lorentz factor depends on the acceleration mechanism and the cooling process of electrons, which is given by $\gamma_{\max}=(6\pi q_{\rm e}\zeta/\sigma_{\rm{T}}B)^{1/2}$, where $q_{\rm e}$ is the charge of electron and the acceleration efficiency $\zeta$ is usually less than unity (Kong et al 2011).

The shocked electrons lose their energies via adiabatic or radiative cooling processes and the evolved distributions of shocked electrons can be obtained by solving the continuity equation, i.e.
\begin{equation}\label{ne0}
\frac{\partial n(\gamma,t)}{\partial t}+\frac{\partial
\dot{\gamma} n(\gamma,t)}{\partial \gamma}=Q(\gamma,t),
\end{equation}
where $\dot{\gamma}={\rm d}\gamma/{\rm d}t$ is the energy loss rate. Under the steady-state assumption $\partial n(\gamma,t)/\partial t = 0$, the solution of the continuity equation can be written as (Khangulyan et al. 2007; Zabalza et al. 2011b; Kong et al. 2012)
\begin{equation}\label{ne}
n(\gamma) = \frac{1}{|\dot{\gamma}|}
\int_{\gamma} Q(\gamma^{\prime}){\rm d}\gamma^{\prime}.
\end{equation}
The electrons with energies above several tens of TeV could be cooled rapidly by the synchrotron radiation, and the distribution index of cooled electrons steepens from $p$ to $p+1$ above the break energy. However, the observed TeV spectrum by \textit{H.E.S.S.} did not show such steepening (Romoli et al. 2017). Similar to van Soelen et al. (2012), we assumed that the shocked electrons are mainly dominated by the adiabatic cooling process, and the shape of electron distribution remains unchanged (Kirk, Ball \& Skj\ae raasen 1999; Zabalza et al. 2011b).

As the energy distribution of shocked electrons is determined, we can calculate the emissivity of the shock by integrating over the distribution of electrons,
\begin{equation}\label{j}
  j(\nu)=\int_{\gamma} n(\gamma)P(\gamma){\rm d}\gamma,
\end{equation}
where $P(\gamma)$ is the sum of synchrotron radiation and anisotropic IC scattering power for an electron with Lorentz factor of $\gamma$ (Rybicki \& Lightman 1979; Aharonian \& Atoyan 1981; Kirk, Ball \& Skj\ae raasen 1999). As the pulsar moves around the companion star, the spatial conditions and the shock parameters are changed accordingly, so the emissivity of the shocked electrons is expected to vary with the orbital phase (Kong et al. 2011).

According to the relativistic hydrodynamical and magnetohydrodynamical simulations by Bogovalov et al. (2008, 2012), the shocked flow is accelerated to very high speed as it propagates far away from the shock apex. Dubus et al. (2010) suggested that the bulk motion of the flows is mildly relativistic, which means that the observed emission from the shock would be boosted or de-boosted at different orbital phases. Taking into account the Doppler effect on the shock radiations, the observed flux on Earth can be calculated as (Dubus, Lamberts \& Fromang 2015)
\begin{equation}\label{F}
  F(\nu)=\frac{1}{d_L^2}\int_V D_{\rm obs}^2j(\nu/D_{\rm obs}){\rm d}V.
\end{equation}
In the calculations, we treat the emitting region as a simple one-zone for simplicity. The Doppler-boosting factor $D_{\rm obs}$ is given by
\begin{equation}\label{Dobs}
  D_{\rm obs}=\frac{1}{\Gamma(1-\beta\cos\theta_{\rm obs})},
\end{equation}
where $\Gamma$ is the bulk Lorentz factor of shocked flows, $\beta=\sqrt{1-\Gamma^{-2}}$. Although the simulations by Bogovalov et al. (2008) suggested that the shocked flows can be accelerated to an extremely high speed ($\Gamma\sim100$), the observed modulations in the X-ray and TeV light curves indicate modest boosting, so we adopt $\beta=1/3$ in the calculations (Dubus et al. 2010). Assuming that the shock has a purely radial structure directing away from the star, the angle between the line of sight (LOS) and the moving direction of shocked flow is given by $\cos\theta_{\rm obs}=\sin{i}\cos(\phi-\phi_{\rm infc})$, where $i$ is the inclination angle of the orbit, $\phi_{\rm infc}=3\pi/2-\omega$ is the true anomaly of the direction of Earth, and $\omega$ is the longitude of periastron.
In this case, the maximum boosting occurs at the inferior conjunction when the flow is moving directly towards the observer (Neronov \& Chernyakova 2008; Dubus et al. 2010; Kong et al. 2012).

\section{Results}

\subsection{Comparisons with observations}
\begin{figure}
  {\includegraphics[width=9cm,height=6cm]{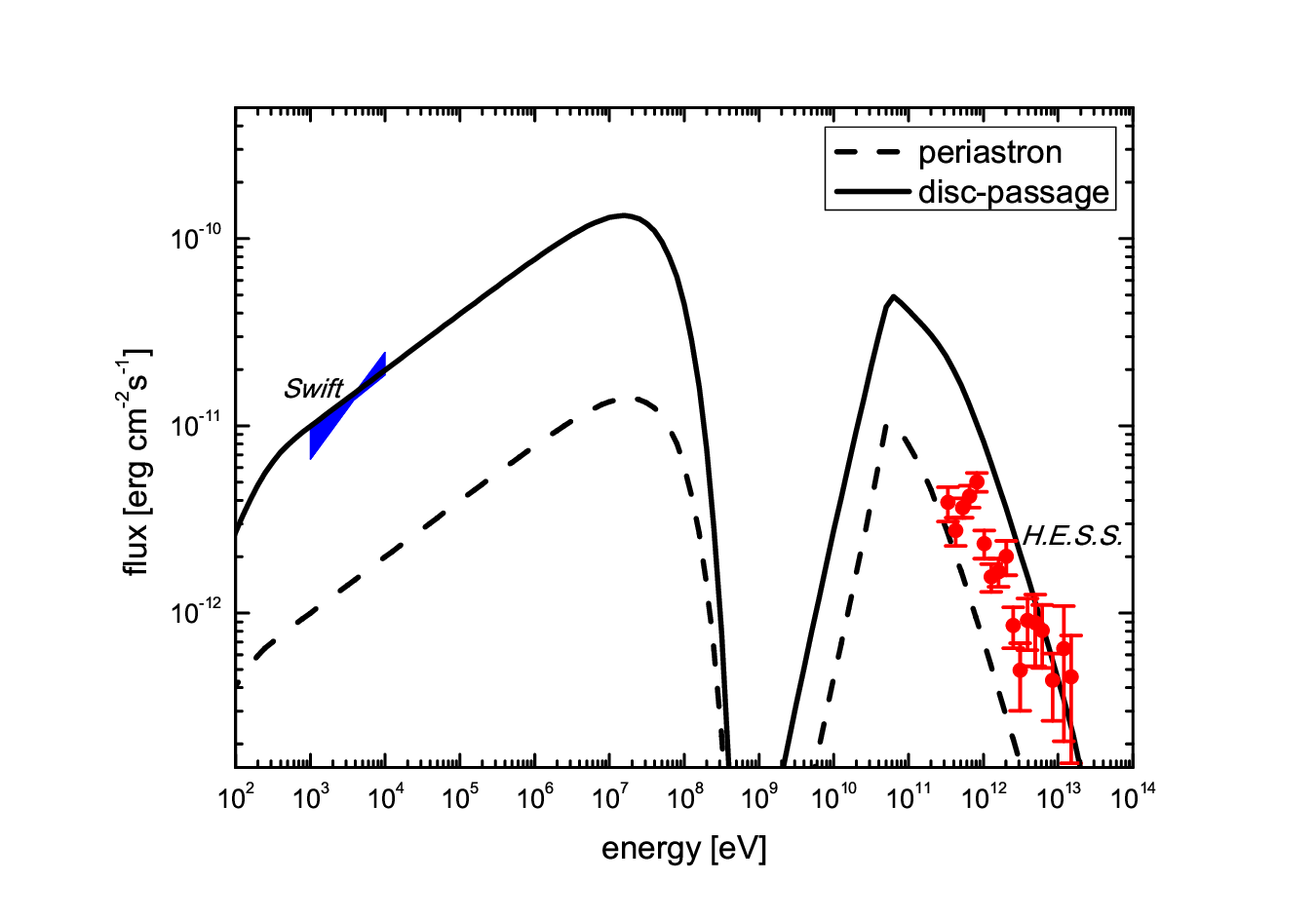}}
  \caption{Calculated spectrum of PSR B1259-63/LS 2883 at periastron (i.e. $\phi=0$, dashed lines) and the disc passage after periastron (i.e. $\phi=\phi_{\rm d}$, solid lines). The \textit{Swift} observational data are taken from Chernyakova et al. (2014) at $23.96$ days after periastron, and the \textit{H.E.S.S.} observational data are taken from the preliminary data reported by Romoli et al. (2017). }
  \label{fig:SED}
\end{figure}

In Fig.\ \ref{fig:SED}, we present the calculated spectrum of synchrotron radiation and IC scattering at periastron ($\phi=0$, dashed lines) and the disc passage after periastron ($\phi=\phi_{\rm d}$, solid lines) with comparisons of observational data.
The \textit{Swift} data are taken from Chernyakova et al. (2014) at $23.96$ days after periastron, and the \textit{H.E.S.S.} data are taken from the preliminary data reported by Romoli et al. (2017).
In our calculations, we assume that the electrons are dominated by adiabatic cooling, so the power index of electrons remains unchanged (Kirk et al. 1999). With $p=2.4$, the resulting photon index of synchrotron radiation in keV X-ray is $\alpha_{\rm X}=(p+1)/2\sim1.7$, and the IC scattering in Klein-Nishina regime in TeV $\gamma$-ray is $\alpha_{\gamma}=(2p+1)/2\sim2.9$, which are consistent with the observational data\footnote{The relation between the  photons index $\alpha$ and the spectral index of electrons $p$ can be derived with the following basic relation: $F(\nu){\rm d}\nu=P(\gamma)n(\gamma){\rm d}\gamma$. For synchrotron radiation and IC scattering in Thompson regime, we have $\nu\propto\gamma^2$ and $P(\gamma)\propto\gamma^2$. Combined with above equations and $n(\gamma)\propto\gamma^{-p}$, we can get $F(\nu)=P(\gamma)n(\gamma){\rm d}\gamma/{\rm d}\nu\propto\nu^{-(p-1)/2}$, so the photon index is $\alpha=(p-1)/2+1=(p+1)/2$; Similarly, for IC scattering in Klein-Nishina regime, we have $\nu\propto\gamma$ and $P(\gamma)\propto\gamma^{1/2}$, so $F(\nu)\propto\nu^{-(2p-1)/2}$, and the photon index is $\alpha=(2p-1)/2+1=(2p+1)/2$. The seed photons for IC scattering in above derivations are assumed to follow a Planck distribution. For simplicity, the effect of $\gamma$-ray absorption on IC spectrum is not considered in this work, which hardens the photon index of $\gamma$-rays.}. We should note that even though the energy density of the photon field is larger than the magnetic energy density in the shock (i.e. $q=u_{\rm ph}/u_{\rm B}\gg1$, see the bottom panel of Fig. \ref{fig:shock}), the synchrotron luminosity could still exceed the IC scattering luminosity because of the suppression by the Klein-Nishina effect (Moderski et al. 2005).

\begin{figure}
  {\includegraphics[width=9cm,height=5cm]{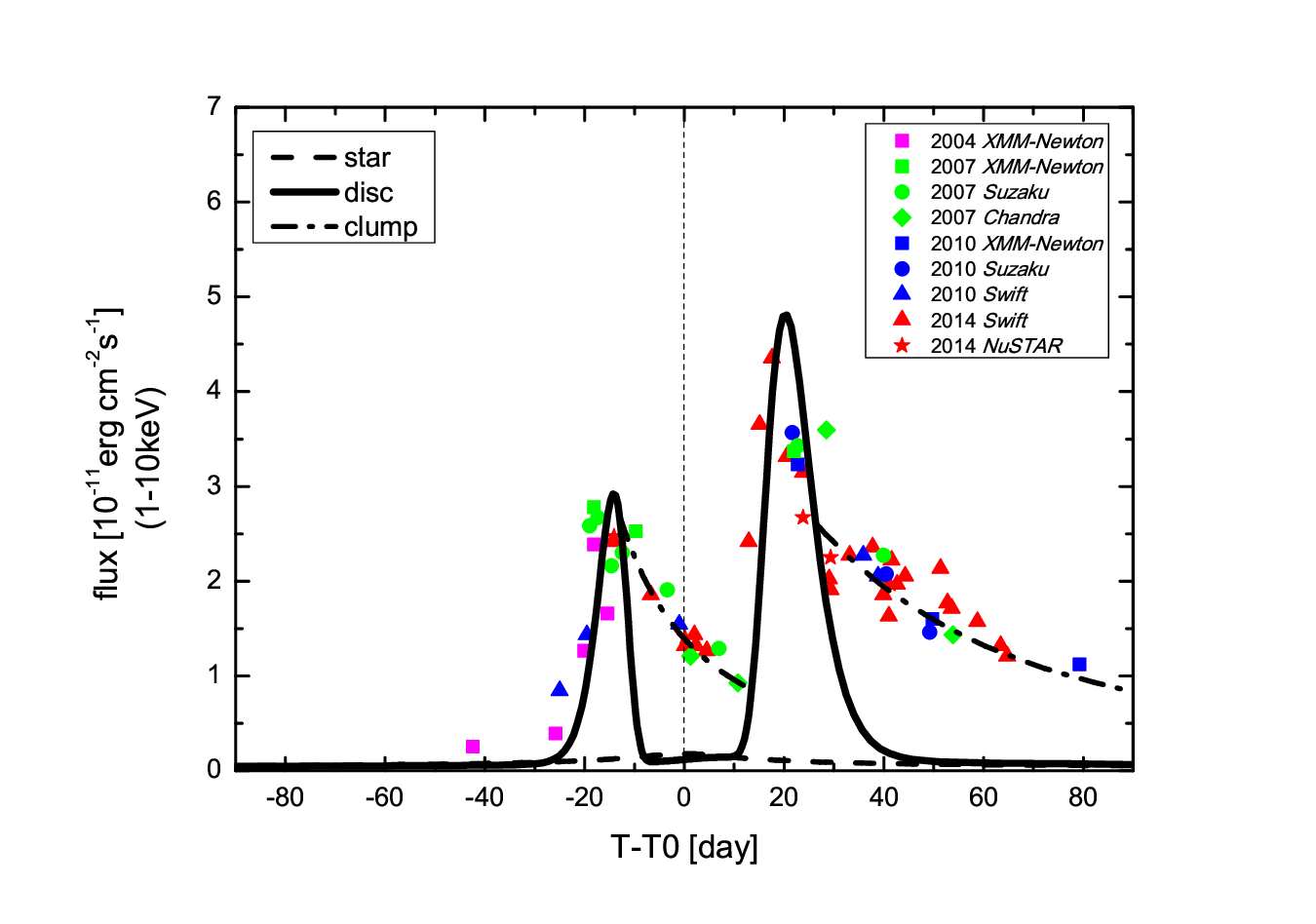}}
  \caption{Integrated X-ray light curves (1-10 keV) with comparisons of observations. The observational data are taken from Chernyakova et al. (2015) without error bars. The dashed line corresponds to the case of synchrotron radiation without the stellar disc while the solid line takes into account the presence of the disc. The dash-dotted line shows the effect of the disc matter clump at the shock front on X-ray emissions. }
  \label{fig:keV}
\end{figure}

The calculated X-ray light curves (1-10 keV) with comparisons of observations are presented in Fig.\ \ref{fig:keV}. The observational data are taken from Chernyakova et al. (2015) without error bars.
The dashed line corresponds to the case without the presence of the stellar disc. As we can see in the figure, the X-ray flux increases slightly as the pulsar is approaching to periastron, since the magnetic field in the shock is highest at periastron without the presence of the disc. The corresponding synchrotron luminosity is determined by the energy density of the magnetic field, which is also highest for the smallest shock distance.
The solid line shows the influence of the stellar disc on the X-ray flux. As we can see that the presence of the stellar disc plays a critical role in producing the double-peak profiles of the X-ray light curve. Because of the compression of the shock by the disc pressure, a higher magnetic field at the shock region during the disc passages results in a stronger synchrotron luminosity and hence produces double-peak profiles as seen in X-ray light curves. As indicated by the observational data, the post-periastron peak is significantly higher (about 1.5 times) than the pre-periastron peak. Since the pulsar is approaching the inferior-conjunction phase during the post-periastron disc passage (see Fig. \ref{fig:binary}), the boosting effect enhances the observed emission, and therefore makes the second peak higher than the first.

\begin{figure}
  {\includegraphics[width=9cm,height=5cm]{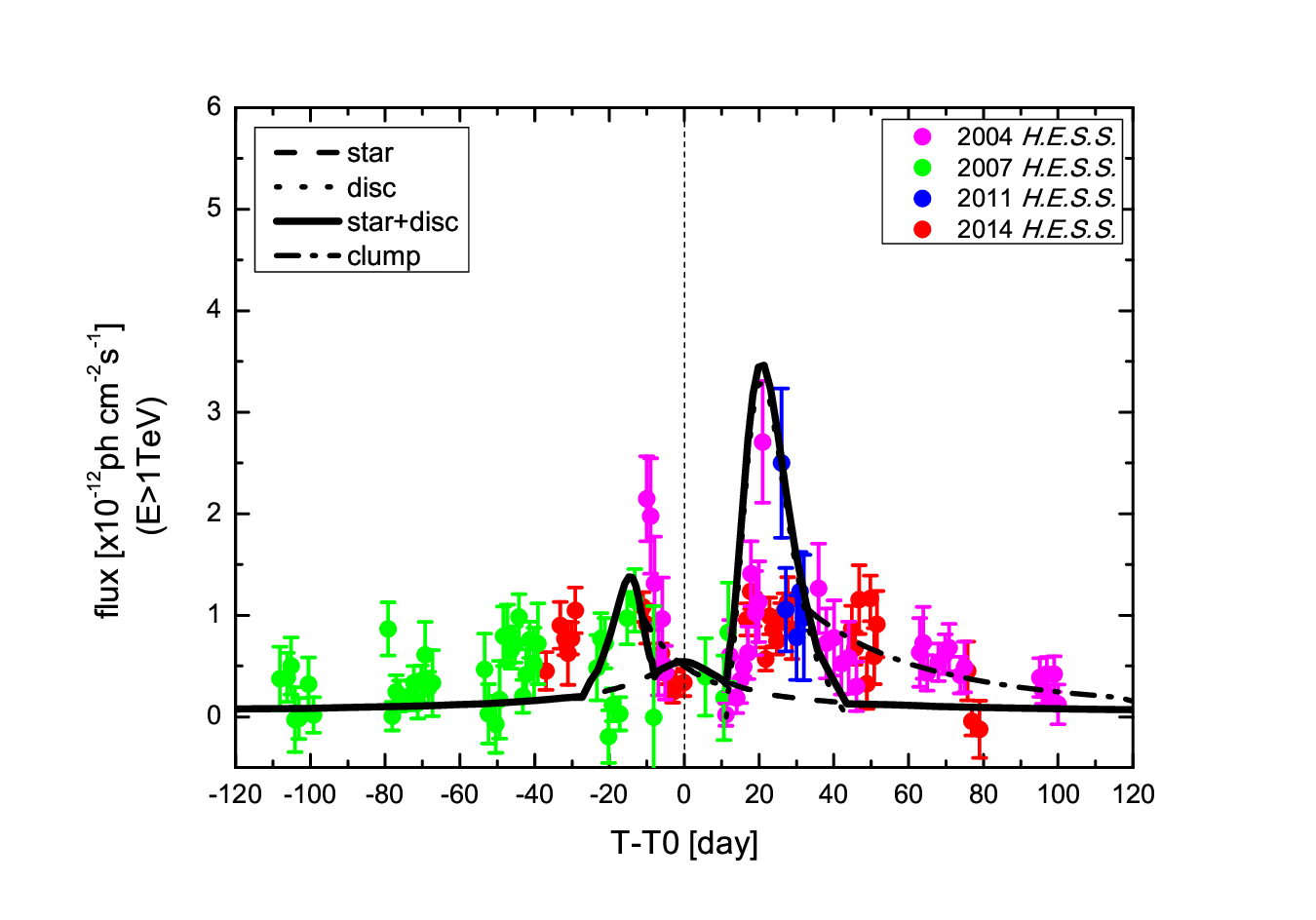}}
  \caption{Integrated TeV light curves ($E>$ 1TeV) with comparisons of observations. The observational data are taken from the preliminary data reported by Romoli et al. (2017). The dashed and dotted lines correspond to the IC scattering with seed photons from the star and the shock heating of the disc. The solid line represents the total flux with both components. The dash-dotted line shows the effect of the disc matter clump at the shock front on TeV emissions. }
  \label{fig:TeV}
\end{figure}

The integrated $\gamma$-ray light curves ($E>1 {\rm TeV}$), compared to observations, are presented in Fig.\ \ref{fig:TeV}. The observational data of \textit{H.E.S.S} are taken from the preliminary data reported by Romoli et al. (2015).
The dashed line corresponds to the TeV light curve produced by IC scattering with the stellar photons. Similar to the X-ray light curve, it shows flux maximum around periastron because the density of the stellar photons at the shock is highest when the pulsar is closest to the star.
The dotted line shows the contribution of the seed photons from the shock heating of stellar disc on $\gamma$-ray flux. As the pulsar passes through the disc region, the density of the soft photons is enhanced by the shock heating process, and the $\gamma$-ray luminosity from IC scattering also increases correspondingly. The solid line shows the total $\gamma$-ray flux with contributions from the star and the shock heating of disc matter. In our calculations, we ignore the possible contribution of the IR emission generated from the disc itself on IC scattering, since its contribution on TeV flux is much smaller than that of stellar photons (see Fig. 8 of van Soelen et al. 2012). Also, we do not consider the $\gamma$-ray absorption on the TeV light curves, since the optical depth for $\gamma$-rays above 1 TeV due to the stellar and disc photons is smaller than unity, and the maxima absorption occurs when the pulsar is around periastron, not the disc passages (Sushch \& van Soelen 2017).

We should note that the X-ray flux exhibits a fast-rise and slow-decay behaviour during the disc passages, and still remains at a high level even after the pulsar leaves the disc region. Tam et al. (2015) found that the declined trend of X-ray light curve can be fitted by an empirical power-law function $F(t)\propto(t-t_{\rm p})^{-0.47}$, where $t_{\rm p}$ is the peak time in light curves. Besides the X-ray light curves, the recent data analysis of the \textit{H.E.S.S.} II also indicated a relatively high TeV flux at several tens of days after periastron (Romoli et al. 2017). The high flux level observed in the X-ray and TeV light curves after the disc passages are somehow strange, since it is generally believed that the enhancements of the X-ray and TeV flux are related to the phases of the pulsar passes through the dense disc environment. This could be because when the pulsar exits from the disc, part of the disc matter is being accumulated at the shock owing to the fast orbital motion of the pulsar, and the shock cannot expand immediately. Similar to Connors et al. (2002), we assume that the shock expands outwards linearly with a velocity of $V$ when it leaves the disc, i.e.
\begin{equation}\label{rx}
  r_{\rm s}(t)\simeq r_{0}+V(t-t_0),
\end{equation}
where $t_0$ and $r_0$ are the time and the shock distance when the pulsar leaves the disc. The expansion velocities adopted in this paper are $V\simeq2\ \rm{km/s}$ and $V\simeq1\ \rm{km/s}$ after the first and second disc passages, which are smaller than those used in Connors et al. (2002). In principal, the expansion velocity is related to the sound speed of disc matter, which is given by $c_{\rm s}=\sqrt{kT/\mu m_{\rm p}}\sim10\ \rm{km/s}$. However, within this expansion velocity, the X-ray and TeV flux decay much faster than the observations, so we suspect the expansion velocity is smaller than $10\ \rm{km/s}$. This could be caused by the fact that the accumulated disc matter is quickly cooled by the expansion process of the shock, and hence the sound speed could be smaller.
When the pulsar leaves the disc region, we assume that the expansion of the shock is an adiabatic process (i.e. $TV^{\gamma-1}=\rm {const}.$, where $\gamma=4/3$ is the adiabatic index), and the temperature evolution of the disc matter clumps can be estimated by
\begin{equation}\label{Tcl}
  T_{\rm clump}\simeq T_{\rm disc}(r_0/r_{\rm s}),
\end{equation}
where $T_{\rm disc}$ is the shock heating temperature of disc matter, and $r_{\rm s}$ is the shock distance given by Eq.(\ref{rx}).
Following the similar description in Section 2, we calculate the expected X-ray and TeV emissions (the dash-dotted lines) due to the disc matter clumps at the shock when the pulsar exits from the disc region in Fig.\ \ref{fig:keV} and Fig.\ \ref{fig:TeV}, respectively.
The declined trend of the light curves based on the model calculation is consistent with the empirical function of Tam et al. (2015).

\subsection{Disc parameters on the shock radiations}
\begin{figure*}
\centering
   \resizebox{16cm}{6.5cm}{\includegraphics{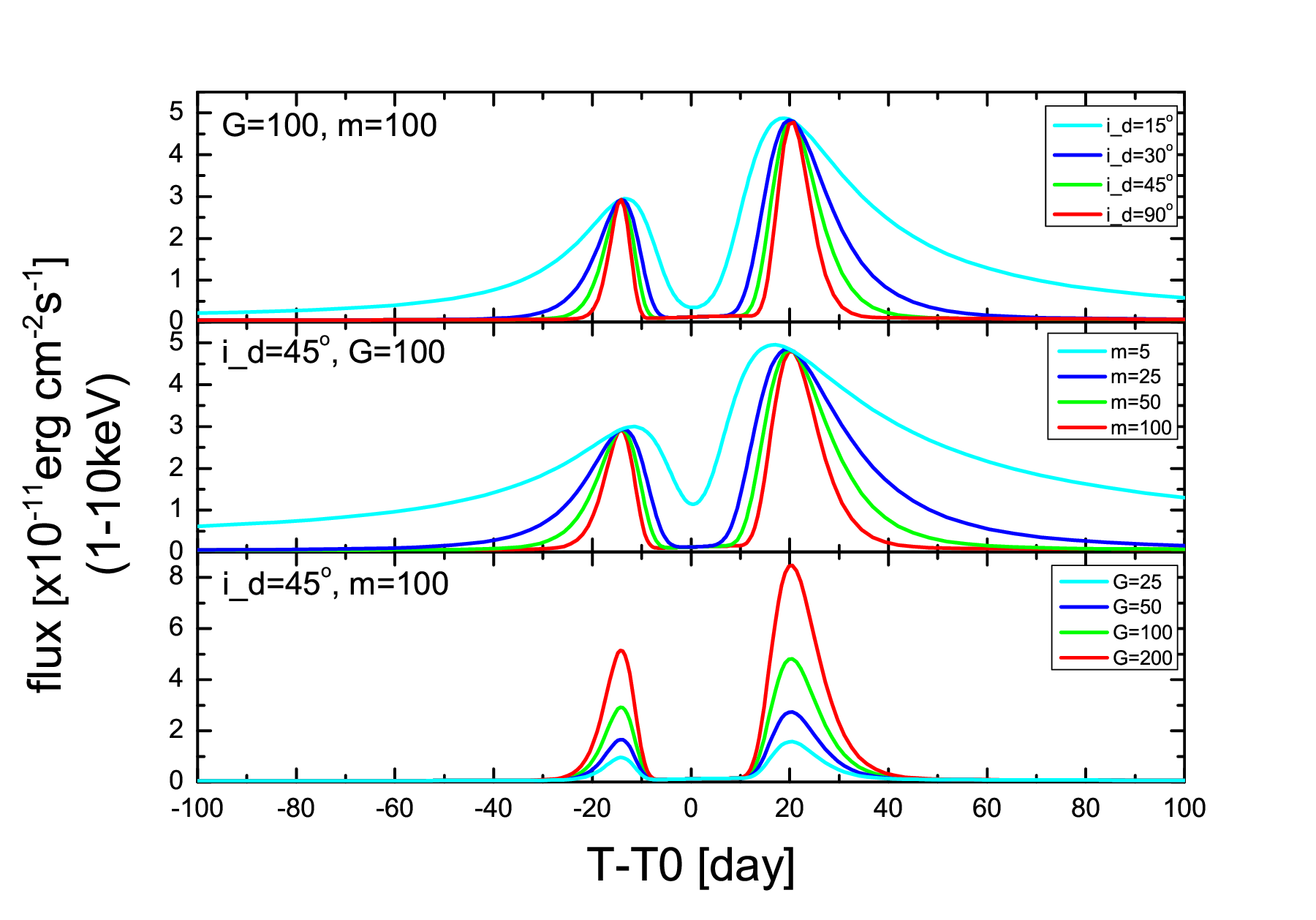}\includegraphics{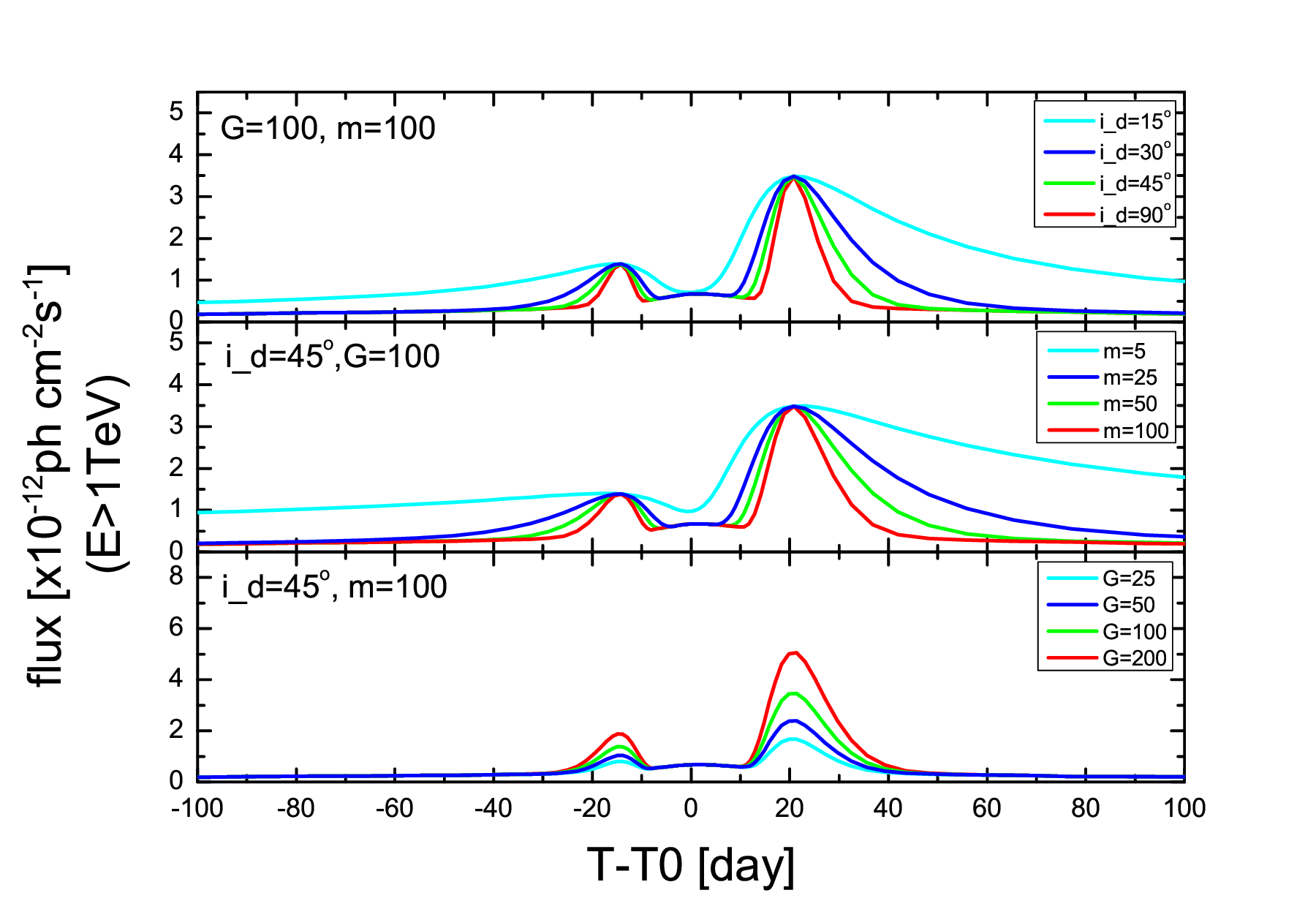}}
     \caption{Disc parameters (i.e. inclination angle $i_{\rm d}$, confinement parameter $m$, and disc-wind pressure contrast $G$) on shock radiations. Left panel: the integrated X-ray light curves in $1-10\ {\rm keV}$ energy band. Right panel: the integrated $\gamma$-ray light curves with $E>1\ {\rm TeV}$ .}
     \label{fig:disc}
\end{figure*}
In previous sections, we adopted a modest disc inclination angle with $i_{\rm d}=45^{\circ}$, and other two disc parameters are set as $G=100$ and $m=100$. Since the presence of the stellar disc has a strong impact on the shock radiation, it is necessary to test other values to see how sensitive the shock emission is to the disc parameters. In Fig.\ \ref{fig:disc}, we calculate the X-ray (left panel) and TeV $\gamma$-ray (right panel) light curves with different values of the inclination angle $i_{\rm d}$, the pressure contrast $G$, and the confinement parameters $m$ of the disc.
We can see that, for a relatively lower value of the inclination angle or the confinement parameters for the disc, the disc opening angle projected on the orbital plane would be larger. For example, in the case of $i_{\rm d}=15^{\circ}$, the stellar disc is almost in the orbital plane, and it means that the pulsar has to take a long time to pass through the disc region.
Alternatively, in the case of $m=5$, with other parameters fixed, the opening angle of the disc is very large, and the pulsar enters the disc region long before periastron, and it is still inside the disc long after periastron. Although a smaller value of $i_{\rm d}$ or $m$ may explain why the X-ray and TeV flux still remains high long after periastron, it is difficult to explain the pulsed radio emission which vanishes around $T_0\pm15$ days, since the optical depth of the pulsed radio signal would be much larger than unity during the disc passages. We note that the pulsed radio flux from the pulsar reappears at $\sim T_0+15$ days (e.g. Chernyakova et al. 2014), which is slightly earlier than the time the pulsar exits from the disc region after periastron in our model. This can be caused by the fact that the powerful wind of the pulsar would create a cavity inside the disc that would allow the observer to receive the pulsed emission from the pulsar magnetosphere without strong absorption of stellar outflows around inferior conjunction (Bosch-Ramon \& Khangulyan 2009).
In the bottom panels of Fig.\ \ref{fig:disc}, we show the X-ray and TeV light curves with different pressure contrasts. As expected, a larger value of $G$ means more significant variations of magnetic field and photon density in the shock during the disc passages and causes a more significant enhancement of the shock emissions.

\section{Discussion and conclusions}
PSR B1259-63/LS 2883 is a unique binary system to study shock physics and constrain the related properties of young pulsars and massive Be stars (e.g. Murata et al. 2003). In this paper, we investigated the influence of the stellar disc on the shock radiations from this system. As the pulsar passes through the dense disc environment, the interaction region of the two winds is compressed into a compact shock.
The additional pressure of the stellar disc pushes the shock surface closer to the pulsar which causes the increase of the magnetic field at the shock. Therefore, the enhancement of the magnetic energy density in the shock during the disc passages increases the synchrotron luminosity, and thus produces the observed double-peak profiles in the X-ray light curve.
As for the TeV $\gamma$-rays, we investigated the IC scattering of shocked electrons with the seed photons from the star and found that the TeV flux is expected to peak around periastron, which is inconsistent with the \textit{H.E.S.S.} observations. The double-peak profiles of the TeV $\gamma$-ray light curve, however, are likely caused by IC process with the seed photons from the shock heating of the stellar disc. As the pulsar passes across the dense disc environment, the pulsar wind sweeps up the outer part of the disc matter into a dense shell. The dense shell absorbs the photons generated by the shocked disc matter and re-emits black-body photons during the disc passages. In this case, the shock heating of the stellar disc provides additional seed photons for IC scattering, and thus produces the double-peak profiles as observed in the TeV $\gamma$-ray light curve. The fitting suggests that the half-opening angle of the disc projected on the orbital plane is $\Delta\phi_{\rm d}\simeq17.25^{\circ}$ with the position of the disc midplane at $\phi_{\rm d}\simeq97.5^{\circ}$. Because of the fast orbital motion of the pulsar around periastron, part of the disc matter might accumulate in the shock front even when the pulsar exits from the disc region, and this might explain why the X-ray and TeV $\gamma$-ray emissions still remain at a high flux level after the disc passages.

The \textit{Fermi}/LAT detections of GeV $\gamma$-rays from PSR B1259-63/LS 2883 during the recent three periastron passages reveal very unexpected results. Especially, its $\gamma$-ray luminosity is of the same order as the spin-down power of the pulsar, which requires an extremely high conversion of spin-down power into $\gamma$-rays (e.g. Johnson et al. 2018). Also, the GeV flare is not accompanied by significant correlations in other wavebands, which means that the GeV emitting particles are distinct from the shocked electrons that radiate keV X-rays and TeV $\gamma$-rays (Dubus 2013). Several models were proposed so far to explain the origin of the GeV flare. Before the detection of GeV emission from PSR B1259-63/LS 2883, Ball \& Kirk (2000) predicted that the optical photons from the companion star are up-scattered to GeV-TeV $\gamma$-rays by the cold pulsar wind electrons, and the $\gamma$-ray emission would reach the flux maximum when the pulsar is around periastron; considering the anisotropic nature of IC, the $\gamma$-rays would peak slightly before periastron. Under the scenario of IC in pulsar wind region, Khangulyan et al. (2011, 2012) considered the time variation of the pulsar wind length towards the observer to explain the GeV flare, and suggested that the sudden increase of $\gamma$-ray luminosity is caused by the rapid increase of the cold pulsar wind region after the pulsar left the disc. The difficulty of their model is the insufficient seed photons from the companion star or the disc for IC scattering to produce such high $\gamma$-rays luminosity during the flare, and additional photon components are required.
Instead, Dubus \& Cerutti (2013) searched for the X-rays from the shock region instead of the photons from the companion star or its disc as seed photons for IC scattering. In this case, the GeV light curve peaks at the phases as the shocked flow crosses through LOS. Similar to Khangulyan et al. (2012), the model requires a very high density of seed photons for the GeV flare, and it was unable to explain the time delay between the X-ray peak after periastron and the GeV flare (Sushch \& B$\ddot{\rm o}$ttcher 2014).
Alternatively, Yi \& Cheng (2017) suggested that a transient accretion disc around the pulsar could be formed after the disc passages and the GeV flare might arise from IC scattering in the cold pulsar wind with the target photons from the transient accretion disc. As the pulsar is passing through the dense equatorial disc, a small fraction of the disc matter might be gravitationally captured by the pulsar. It takes a few days to form the accretion disc and this can explain the time delay of the GeV flare with the disc passages. However, the recent timing observation of PSR B1259-63 shows no evidence of a transient accretion disc (Shannon et al. 2014;  Yi \& Cheng 2018).

Producing the observed GeV $\gamma$-ray luminosity with the order of spin-down luminosity of the pulsar $L_{\gamma}\sim L_{\rm sd}$ requires a high photon density for IC scattering that the star or the disc cannot provide during the flare (Johnson et al. 2018). In this paper, we discussed the case that part of the disc matter is accumulated in the shock when the pulsar exits from the disc region. Since it is suggested that the GeV emission is likely produced by IC scattering in the cold pulsar wind region, the optical/IR photons from the disc matter clumps can supply an additional seed photon field even after the pulsar leaves the disc. We expect that the cold pulsar wind up-scattering of the photons from the disc matter accumulated at the shock, and possibly with the contribution of boosted synchrotron emission from the shock, might explain the mysterious GeV flare. A detailed calculation will be done in the subsequent study.

There are some other high-mass gamma-ray binaries showing many similarities to PSR B1259-63/LS 2883, such as PSR J2032+4127/MT91 213 (Lyne et al. 2015), HESS J0632+057 (Hinton et al. 2009), and LS I+61$^{\circ}$303 (Albert et al. 2006). All of these binaries contain a compact object that is orbiting around a Be star.
Especially, PSR J2032+4127/MT91 213 is the second  high-mass gamma-ray binaries whose compact object was identified as a pulsar. Although the compact objects in the latter two binaries are not confirmed yet, the absence of accretion signals supports a rotation-powered pulsar origin of the compact object. The companion stars of these systems are type Be stars, which possess a dense equatorial disc; the orbital variations of the X-ray/TeV $\gamma$-ray light curves are generally believed to related to the disc passages, which are similar to the case of PSR B1259-63/LS 2883. Since the presence of the equatorial disc will have a strong impact on non-thermal emissions, a detailed investigation of multiwavelength emissions from these binaries could provide us with a better understanding of shock physics and the geometries of gamma-ray binaries.

\begin{acknowledgements}
We are very grateful for the valuable suggestions and comments of the referee, which improved the manuscript significantly. J.T is supported by the National Science Foundation of China under 11573010, 11661161010, U1631103 and U1838102. Y.W.Y is supported by the National Natural Science Foundation of China under 11822302 and 11833003. K.S.C is supported by GRF grant under 17302315.
\end{acknowledgements}

\end{document}